
\input harvmac
\def\np#1#2#3{Nucl. Phys. B{#1} (#2) #3}
\def\pl#1#2#3{Phys. Lett. {#1}B (#2) #3}

\def\prep#1#2#3{Phys. Rep. {#1} (#2) #3}

\def\ev#1{\langle#1\rangle}
\def\pf{{\rm Pf ~}}
\def\tilde{\widetilde}
\def\NNN{\cal{ N}}

\def\psqr#1#2{{\vcenter{\vbox{\hrule height.#2pt
	\hbox{\vrule width.#2pt height#1pt \kern#1pt
	\vrule width.#2pt}
	\hrule height.#2pt \hrule height.#2pt
	\hbox{\vrule width.#2pt height#1pt \kern#1pt
	\vrule width.#2pt}
	\hrule height.#2pt}}}}
\def\sqr#1#2{{\vcenter{\vbox{\hrule height.#2pt
	\hbox{\vrule width.#2pt height#1pt \kern#1pt
	\vrule width.#2pt}
	\hrule height.#2pt}}}}
\def\square{\mathchoice\sqr65\sqr65\sqr{2.1}3\sqr{1.5}3}
\def\doub{\mathchoice\psqr65\psqr65\psqr{2.1}3\psqr{1.5}3}

\Title{hep-th/9510148, RU-95-66}
{\vbox{\centerline{Duality in SUSY
$SU(N)$ with an Antisymmetric Tensor}}}
\bigskip
\centerline{P. Pouliot}
\vglue .5cm
\centerline{Department of Physics and Astronomy}
\centerline{Rutgers University}
\centerline{Piscataway, NJ 08855-0849, USA}

\bigskip

\noindent

We present a dual description for $SU(N)$
supersymmetric gauge theory with an antisymmetric
tensor and fundamentals, and no superpotential.
This duality is derived {}from the dualities of Seiberg.
Under a perturbation of the superpotential,
the dual theory breaks supersymmetry at tree level.

\Date{10/95}

\nref\sem{N. Seiberg, \np{435}{1995}{129}}%
\nref\powerd{N. Seiberg, the Power of Holomorphy -- Exact
Results in 4D SUSY Field Theories, Proc. of PASCOS 94, hep-th/9408013,
 RU-94-64, IASSNS-HEP-94/57; the Power of Duality
 -- Exact Results in 4D SUSY Field Theories,
Proc. of PASCOS 95 and Proc. of the
Oskar Klein Lectures, hep-th/9506077, RU-95-37, IASSNS-HEP-95/46;
K. Intriligator and N. Seiberg,
hep-th/9509066, RU-95-48, IASSNS-HEP-95/70}
\nref\sv{M.A. Shifman and A. I. Vainshtein, \np{277}{1986}{456};
\np{359}{1991}{571}}
\nref\ads{I. Affleck, M. Dine and N. Seiberg,
\np{241}{1984}{493}; \np{256}{1985}{557}}%
\nref\cernrev{D. Amati,
K. Konishi, Y. Meurice, G.C. Rossi and G.
Veneziano, \prep{162}{1988}{169} }
\nref\olive{C. Montonen and D. Olive, \pl{72}{1977}{117}}
\nref\switten{N. Seiberg and E. Witten, \np{426}{1994}{19};
 \np{431}{1994}{484}}

\nref\son{K. Intriligator and N. Seiberg,
\np{444}{1995}{125}; hep-th/9506084,
RU-95-40, IASSNS-HEP-95/48}
\nref\intpou{K. Intriligator and P. Pouliot, \pl{353}{1995}{471}}
\nref\pouliot{P. Pouliot, \pl{359}{1995}{108}}
\nref\poustr{P. Pouliot and M. Strassler,
RU-95-67, to appear}

\nref\kut{D. Kutasov, \pl{351}{1995}{230}}
\nref\ofer{O. Aharony, J. Sonnenschein and S. Yankielowicz,
hep-th/9504113, TAUP-2246-95, CERN-TH/95-91}
\nref\kuschwim{ D. Kutasov and A. Schwimmer,
\pl{354}{1995}{315}}
\nref\berk{M. Berkooz, hep-th/9505067, RU-95-29}
\nref\intril{K. Intriligator, \np{448}{1995}{187}}
\nref\leighstr{R.G. Leigh
and M.J. Strassler, \pl{356}{1995}{492}}
\nref\ilstr{K. Intriligator, R.G. Leigh and M.J. Strassler,
hep-th/9506148, RU-95-38}
\nref\mur{H. Murayama, \pl{355}{1995}{187}}
\nref\pop{E. Poppitz and S. P. Trivedi, hep-th/9507169,
EFI-95-44}
\nref\dnns{M. Dine, A. Nelson, Y. Nir and Y. Shirman,
hep-ph/9507378, SCIPP 95/32,
UW-PT/95-08, WIS-95/29/Jul-PH}
\nref\ithomas{K. Intriligator and S. Thomas,
SLAC-PUB-95-7041, to appear}

One year ago, a remarkable feature of gauge field theories
was discovered \sem. Certain $\NNN$=1 supersymmetric field
theories were shown by Seiberg to have a
completely equivalent description
at large distances in terms of a different
SUSY gauge theory, with a different gauge group and
matter content. For reviews and lists of references,
see \powerd. For earlier work on these SUSY theories
see \refs{\sv- \cernrev}.
This duality is a generalization of the one of Montonen and
Olive of $\NNN$=4 \olive\ and some
$\NNN$=2 SUSY gauge theories \switten.

The original examples in \sem\ included $SU(N)$, $SO(N)$ and
$Sp(N)$ gauge theories with fundamentals. They were further
studied in \refs{\son,\intpou }. It is not straightforward
to find dual descriptions for other gauge groups and matter content.
A dual description for SUSY $G_2$ gauge theory with fundamentals
was found in \pouliot. It was extended to
other theories in \poustr.
A wealth of partial results on dualizing
many other theories were found in \refs{\kut - \ilstr},
in an approach
which requires
to perturb the theory by a superpotential in order to
perform the duality.

In this letter,
the theory that we consider is an $SU(N)$
supersymmetric gauge theory without a superpotential.
The matter fields are chiral superfields
transforming in one antisymmetric
tensor
representation
$A$, $F$ fundamentals $Q$ and $N+F-4$
antifundamentals $\overline{Q}$.
That is, there is $F$ extra flavors of $Q$ and $\overline Q$
beyond the necessary antifundamentals to cancel the $SU(N)^3$ anomaly.
We present a dual description for this model
for $N$ odd, which we refer to
as the electric theory.
A dual for this model with a superpotential was
obtained in \berk\ for $N$ even.
We derive this duality {}from
the elementary dualities of Seiberg following the
idea of \berk\ of {\sl deconfining} the antisymmetric tensor,
by introducing an auxiliary gauge group.
We also propose a dual description for $SU(N)$ with
an antisymmetric tensor and a conjugate antisymmetric tensor,
fundamentals and antifundamentals, which we do not analyze in detail.
We then study some of the features of the model
with one antisymmetric tensor in detail.
One feature is that a baryon operator is
an elementary field in the dual description.
Also, this duality flows to the $SU$ and $Sp$ dualities
of \sem\ and that of \berk.
Another interesting feature is that this model
was shown in \ads\ to break supersymmetry dynamically.
This aspect was further studied in \refs{\mur - \ithomas}.
The novelty here is that SUSY breaking is studied
at weak coupling in the dual description.

We first describe the electric theory.
Under the continuous non-anomalous
$SU(F)\times SU(N+F-4)\times U(1)_1\times U(1)_2\times U(1)_R$
symmetries, the fields transform as:
\eqn\electric{\matrix{
& SU(N) &  SU(F) & SU(N+F-4) & U(1)_1 & U(1)_2 & U(1)_R \cr
\cr
A &  \doub & 1 & 1 & 0 & -2F & {-12\over N} \cr
Q & \square
 & \square & 1 & 1 & N-F & 2 -{6\over N} \cr
\overline{Q} & \overline{\square} & 1 & \square
& {-F\over N+F-4} & F & {6\over N }. \cr } }
There is a two parameter family of R-symmetries;
our choice of $U(1)_R$ is for convenience.

The flat directions can be conveniently described by
the following gauge invariant chiral operators: mesons
$M\equiv Q\overline{Q}$ and $H\equiv A\overline{QQ}$;
baryons
$\overline{B}\equiv \overline{Q}^N$ ($F\ge 4$) and
$B_{k}\equiv Q^{k} A^{N-k\over 2}$ ($N$ and $k$ both odd or
both even, $k\le \min(N,F)$).
These operators are not all independent \pop,
but classically constrained.
We will not discuss these constraints.
When $M$ gets an expectation value of rank $r$,
the theory is higgsed to $SU(N-r)$ with,
as the remaining matter content,
an antisymmetric tensor $\hat A$,
$N-r+F-4$ antifundamentals $\hat{\overline{Q}}$ and still
$F$ fundamentals $\hat Q$ (among which $r$ are coming {}from
$A$). Similarly,
when $H$ gets an expectation value of
rank $2r$,
the group is higgsed to $SU(N-2r)$ with
an antisymmetric tensor $\hat A$, $N-2r+F-4$
antifundamentals $\hat{\overline{Q}}$
and
$F$ fundamentals $\hat Q$ remaining.
When $B_{k}$ gets an expectation value,
the theory is higgsed to $Sp({N-k\over 2})$ with
 $N-k+2F-4$ fundamentals $\hat Q$ remaining,
coming {}from $\overline Q$ and $Q$.
When $\overline{B}$ gets an expectation value, the theory
is completely higgsed.

For completeness, we
 briefly summarize the results of \pop\ in our notation.
For $F\ge 3$, by holomorphy, the symmetries and weak coupling,
no superpotential can be generated dynamically.
For $F=0$ and $N$ odd, there is no invariant that can appear
in the superpotential,
which remains $W=0$;
upon adding a tree level term $\lambda H$
to the superpotential,
it was shown that the theory has no supersymmetric
vacuum \refs{\ads, \mur,\pop}.
For $F=0$ and $N$ even, a superpotential
$(\Lambda^{2N+3}/(B_0\pf H))^{1/3}$ is generated by gluino condensation
in an $Sp(2)$ subgroup of $SU(N)$.
For $F=1$, a superpotential is generated by gluino condensation
in an $Sp(1)$ subgroup of $SU(N)$;
it is $(\Lambda^{2N+2}/ (B_1 \pf H))^{1/2}$ (for $N$ odd) and
$(\Lambda^{2N+2}/(B_0 M H^{(N-4)/2}))^{1/2}$ (for $N$ even).
For $F=2$, a superpotential is generated by instantons;
it is $\Lambda^{2N+1}/(B_1 M H^{(N-3)/2})$ (for $N$ odd)
and $\Lambda^{2N+1}/(B_0M^2H^{-1}\pf H+B_2\pf H)$ (for $N$ even).
For $F=3$, the singular classical
moduli space is smoothed out quantum mechanically,
as the classical constraint is modified to
$B_1 M^2 H^{-1}\pf H-B_3 \pf H = \Lambda^{2N}$ (for $N$ odd) and
$B_0 M^3 H^{(N-4)/2} + B_2 M H^{(N-2)/2}=\Lambda^{2N}$ (for $N$ even).
For $F\ge 4$, the classical and the quantum moduli spaces of vacua
are the same.
For $F=4$, the theory is one of massless mesons and baryons at
the origin of the moduli space. It consists of
the independent gauge invariant degrees of freedom
$M$, $H$, $B_1$, $B_3$ and
$\overline{B}$ (for $N$ odd) and of
$M$, $H$, $B_0$, $B_2$, $B_4$ and
$\overline{B}$ (for $N$ even). This satisfies the 't Hooft anomaly matching
conditions. Their interaction can be described by the
confining superpotentials
$(B_1M^3H^{(N-3)/2}+ B_3 M H^{(N-1)/2}+\overline{B}B_1B_3)
/\Lambda^{2N-1}$ (for $N$ odd) and
$(B_0 M^4 H^{-2}\pf H + B_2 M^2 H^{-1}
\pf H + B_4 \pf H + \overline{B}B_0B_4
+\overline{B} B_2^2)/\Lambda^{2N-1}$ (for $N$ even),
and the resulting equations of motion yield constraints on the
expectation values of the fields.

When $F\ge 5$, the theory at the origin
of the moduli space is in a non-Abelian
Coulomb phase.
The main result of this paper
is that for $N$ odd,
the $SU(N)$ theory described above,
which we will call electric, has a dual description
in terms of an
$SU(F-3)\times Sp(F-4)$ gauge theory with
five species of dual quark superfields:
a field $x$ transforming as a fundamental under both gauge
groups, a
conjugate antisymmetric tensor $\overline{a}$, a fundamental $p$ and
$F$ antifundamentals $\overline{q}$ of $SU(F-3)$
and also $N+F-4$ fundamentals $l$ of $Sp(F-4)$.
Furthermore, this dual magnetic theory, later
referred to as the first dual, contains elementary gauge
singlet fields $M$, $H$ and $B_1$.
The global non-anomalous symmetry is the same as in the electric
theory and the transformation
properties of these fields are listed below.
\eqn\magnetic{\matrix{
& \scriptstyle{SU(F-3)} & \scriptstyle{Sp(F-4)}
 & \scriptstyle{SU(F)} & \scriptstyle{SU(N+F-4)}
 & U(1)_1 & U(1)_2 &U(1)_R \cr
\cr
x  & \square & \square & 1 & 1 & {-F\over F-3} & 0 & -1 \cr
p & \square  & 1 & 1 & 1  & {-F\over F-3} & NF & 6 \cr
\overline{a} & \overline{\doub} & 1 & 1
& 1 & {2F\over F-3} & 0 & 4 \cr
\overline{q}  & \overline{\square} & 1  & \overline{\square} & 1
& {3\over F-3}
& -N & 0 \cr
l & 1 & \square & 1 & \overline{\square} & {F\over N+F-4} & 0 & 1 \cr
\cr
M & 1 & 1 & \square & \square
& {N-4\over N+F-4}
& N & 2 \cr
H & 1 & 1& 1 & \doub & {-2F\over N+F-4} & 0 & 0 \cr
B_1 & 1 & 1  & \square & 1
& 1 & N(1-F) & -4. \cr} }
This magnetic theory has the superpotential
\eqn\magsuper{
W= M\overline{q} l x + H l l  + B_1 p\overline{q} + \overline{a}x^2.}
It is the most general superpotential allowed by the symmetries,
holomorphy, and smoothness near the origin in field space.
Note that the form of $W$, along with the identification of
the operators $M$, $H$ and $B_1$ of the dual with those of the
electric theory, the $U(1)$ charges are determined.
Under this charge assignment, the three $U(1)$s are anomaly free
in both the electric and magnetic theory. The 't Hooft anomaly
matching conditions for the full $SU(F)\times SU(N+F-4)\times
U(1)_1\times U(1)_2\times U(1)_R$
global symmetry are satisfied.
These conditions are clearly satisfied for any complex numbers
$F$ and $N$.
But the magnetic theory presented here is
certainly not valid for $N$ even
(see the mapping of operators below) and therefore more checks
are required to establish
the duality for $N$ odd. One route is to check the maps
of operators, perturbations of the superpotential,
flat directions, etc. We will return to this briefly later.
Alternatively, this duality can be
derived {}from the elementary dualities of Seiberg.

To this end,
consider an $Sp({N-3\over 2})$ SUSY gauge theory (for $N$ odd)
with $N+1$ fundamentals, $y_i$ and $z$,
and $N$ singlets $\overline{P}^i$, $i=1,\ldots,N$.
This theory confines \intpou\ and yields a superpotential
$W=y^N z= A^{N-1\over 2} P$ for the gauge invariant fields
$A_{ij}\equiv  y_i y_j$ and $P_i\equiv zy_i$. Add
to the superpotential a coupling
$zy_i \overline{P}^i$, which is a mass term for $P_i$ and $\overline{P}^i$.
This coupling breaks the $SU(N+1)\times SU(N)$ flavor symmetry to
$SU(N)\times U(1)$.
Integrating out $P_i$ and $\overline{P}^i$, we get a theory with
${N(N-1)\over 2}$ singlets $A$ and no superpotential;
there is no constraint on the light fields $A$.
Now consider gauging the $SU(N)$ flavor symmetry, under which
$A$ is an antisymmetric tensor; introduce more $Sp({N-3\over 2})$ singlets
$Q$ and $\overline{Q}$, which are fundamentals and antifundamentals of
$SU(N)$, to cancel the $SU(N)^3$ anomaly. Therefore,
the $SU(N)\times Sp({N-3\over 2})$ expanded theory
just described is equivalent
to the electric theory \electric. The charge
assignments in the expanded theory follow {}from the
relations
$W=y^N z+zy\overline{P}$,
$A_{ij}\equiv  y_i y_j$ and $P_i\equiv zy_i$, to obtain:
\eqn\expanded{\matrix{
& SU(N) & Sp({N-3\over 2})
 &  SU(F) & SU(N+F-4) & U(1)_1 & U(1)_2 & U(1)_R \cr
\cr
y & \square & \square & 1 & 1 & 0 & -F & {-6\over N} \cr
z & 1 & \square & 1 & 1 & 0 & FN & 8 \cr
\overline{P} & \overline{\square} & 1 & 1
& 1 & 0 & F(1-N) & -6+{6\over N} \cr
Q & \square
 & 1 & \square  & 1 & 1 & N-F & 2 -{6\over N} \cr
\overline{Q} & \overline{\square} & 1 & 1 & \square
& {-F\over N+F-4} & F & {6\over N }. \cr } }
The $SU(N)$ gauge group has $N+F-3$ flavors, and is in a non-Abelian
Coulomb phase precisely for $F\ge 5$; it can be dualized by the $SU(N)$
duality of \sem. The result is a dual description
in terms of an $SU(F-3)\times Sp({N-3\over 2})$ gauge
theory with the following
matter content:
\eqn\magnetictwo{\matrix{
& SU(F-3) & Sp({N-3\over 2}) & SU(F) & SU(N+F-4)  \cr
\cr
\overline{x} & \overline{\square} & \square  & 1 & 1 \cr
p  & \square & 1 & 1 & 1  \cr
\overline{q}  & \overline{\square} & 1 & \overline{\square} & 1 \cr
q & \square & 1 & 1 & \overline{\square} \cr
\overline{l} & 1 & \square  & 1 & \square \cr
\cr
M=Q\overline{Q} & 1 & 1 & \square & \square  \cr
B_1=QA^{N-1\over 2} & 1 & 1 & \square & 1. \cr} }
The superpotential is
\eqn\magsuper{
W= Mq\overline{q} + B_1 p\overline{q} + \overline{l} \overline{x}q.}
Now observe that the $Sp({N-3\over 2})$ gauge group has
$N-7+2F$ fundamentals, so that precisely for $F\ge 5$, it
is in a non-Abelian Coulomb phase and can be dualized by
the $Sp$ duality of \sem. The result
after integrating out the massive fields
is the dual description
presented above \magnetic.

Going back to the first dual, note
that its $SU(F-3)$ gauge group
has matter content in an antisymmetric tensor, fundamental
and antifundamental representations, and thus can be dualized with
the duality presented in this letter; this procedure can be iterated
$n$ times to yield dual descriptions with gauge group
$SU(F-3)\times \prod^{n}_{i=0} Sp(F-4)_i$.
One generalization of this construction is to
dualize $SU(N)$ (for $N$ odd)
with
an antisymmetric tensor $\doub$,
a conjugate antisymmetric tensor
$\overline{\doub}$ and $F$ flavors of
fundamentals $\square$ and antifundamentals $\overline{\square}$,
without a superpotential.
This model with a superpotential was studied in \ilstr.
We suggest that the iteration procedure
just mentioned could account
for the multiple vacua obtained in \ilstr\
of the form $SU(n_0)\times \prod_i Sp(n_i)$.

We now study the features of the $SU(N)$ model with
only one antisymmetric tensor in more detail.
The baryon operators of the electric theory are mapped to
baryons of the
magnetic theory in the following way:
\eqn\mapping{
\overline{Q}^{N} \to p (x l)^{F-4} \qquad
Q^{k} A^{N-k\over 2} \to \tilde B_{F-k}\equiv \overline{q}^{F-k}
\overline{a}^{k-3\over 2}, }
(for $k=3,5,\ldots,\min(N,F)$).
Under this mapping, all the continuous
global symmetries are preserved.
Note however that the baryon $B_1$ is an elementary field
in the dual description.

Consider giving an expectation value of rank one to $M$;
for definiteness $\ev{M_{F,N+F-4}}\neq 0$. The electric theory
is higgsed to $SU(N-1)$. The result is a dual description for
$SU(N_c)$ with an antisymmetric tensor and an even number of
colors $N_c$.
Then add a term proportional to
$B_{1,F}\equiv \pf \hat A$ to the superpotential of the electric theory.
The effect on these perturbations on the dual is more easily seen on
the second dual \magnetictwo, for which the superpotential is
cubic. The effect is to give a mass to
$q_{N+F-4}$ and $\overline{q}_F$ and
to higgs the $SU(F-3)$ part of the gauge group to $SU(F-4)$
by giving an expectation value
$\ev{p\overline{q}_F}\neq 0$. The fields $B_1$, $M_{F,i}$,
$M_{j,N+F-4}$, $p$, $\overline{q}_F$ are massive.
After integrating them out,
one recovers precisely the first of Berkooz' duals \berk.
His other dual is obtained by dualizing the $Sp({N-3\over 2})$
gauge group.

It is straightforward to check that the
duality between the theories
\electric\ and \magnetic\ reduces to the duality of \sem\ for $N=3$.
Since the antisymmetric tensor $A$
is a $\overline{3}$ of $SU(3)$, we have a larger flavor symmetry
$SU(F)\times SU(F)$, with $F$ flavors of $3$ and
$\overline{3}$. In the dual \magnetic,
the $Sp(F-4)$ gauge group has $F-2$ flavors and thus confines \intpou,
 yielding operators
$\overline{H}\equiv l^2$, $a\equiv x^2$,
$q\equiv xl$ and a confining superpotential
$\sum_{n=0} a^n \overline{H}^{n+1} q^{F-3-2n}$.
The superpotential of the magnetic theory has also the
piece $Mq\overline{q}+H\overline{ H}+B_1 p\overline{q}+
a\overline{a}$.
The fields $a$, $\overline{a}$, $H$, $\overline{H}$ are all massive,
and after they are integrated out, there
remains only $Mq\overline{q}+B_1 p\overline{q}=
 \hat M\hat q \hat{\overline{q}}$
for $\hat q =(q,p)$, $\hat{\overline{q}}=\overline{q}$,
$\hat M=(M,B_1)$,
which is precisely the $SU(F-3)$ dual of $SU(3)$ with $F$ flavors
of \sem\ with the correct superpotential.

Another special case is for $F=5$.
Then the dual is $SU(2)\times SU(2)$ and
the field $\overline{a}$ is just a singlet
that is identified with
the baryon $B_5$ of the electric theory.

We now analyze the baryonic flat directions $\ev{B_k}\neq 0$.
The electric theory becomes $Sp({N-k\over 2})$ with $N-k+2F-4$
fundamentals as mentioned above.
When $k=1$ and $F> 5$, the effect on the magnetic description
is to provide a new dual
(though possibly a very strongly coupled one)
for $Sp$ gauge theories with fundamentals.
For $k>1$,
$\ev{\tilde B_{F-k}}=
\ev{\overline{q}^{F-k}\overline{a}^{k-3\over 2}}\neq 0$ in the dual.
The $SU(F-3)$ group is higgsed to $Sp({k-3\over 2})$ with
$k+1$ fundamentals $\overline{q}$ and $p$ remaining since
the fundamentals $x$ are made massive by the
$\overline{a}xx$ superpotential.
This $Sp({k-3\over 2})$ gauge group confines, yielding a confining
superpotential $\overline{q}^kp$ and the
fields $\overline{B}_1=\overline{q}p$
and $B_1$ are massive. The massless fields that remain
and that transform under the $Sp(F-4)$ gauge group
are $F-k$ fundamentals $x$ and $N+F-4$ fundamentals $l$ of $Sp(F-4)$.
After integrating out the massive fields,
the superpotential for the light
fields
is $Mxl+Hll+\overline{a}xx$ where an
expectation value $\ev{\overline{q}}$ has
been absorbed in a redefinition of $M$. Rewriting
$\hat M=\pmatrix{ H & M \cr
-M & \overline{a} \cr}$ and $\hat q=(l,x)$, we see that the
theories are precisely dual under the $Sp(N_c)$ duality of
\sem. When $F=5$ and $k=1$,
after a similar analysis, one recovers
the $Sp$ duality of \sem.
At first sight, this magnetic theory for $F> N$ seems to have
more operators than the electric theory, namely the
$\tilde B_{F-k}$ for $k>N$. However, whenever
these operators are non-vanishing,
there is no supersymmetric vacuum: the $Sp(F-4)$ theory generates
dynamically for $\hat q$
a superpotential which slopes to infinity for $k> N+2$, or a constraint
$\pf \hat q=\Lambda^{2(F-3)}$ for $k=N+2$; but the
equations of motion cannot be satisfied since the equation of motion for
$\hat M$ sets $\hat q\hat q$ to zero.
Therefore the magnetic theory has the same chiral ring as
the electric theory.

Consider giving an expectation value to $\overline{B}$.
The electric theory is completely higgsed. There
remains $NF$ fields $Q$, ${N(N-1)\over 2}$ fields $A$ and
$N(F-4)$ fields $\overline{Q}$ as massless singlets.
In the magnetic theory $\ev{p(xl)^{F-4}}\neq 0$. The
$SU(F-3)\times Sp(F-4)$ group is completely higgsed.
The fields $p$, $x$ and $(F-4)^2$ of the fields
$l$ are eaten.  The superpotential makes
$B_1$, $\overline{q}$, $\overline{a}$ as well as
some components of $l$, $M$ and $H$ massive.
The massless components of $l$, $M$ and $H$
that remain correspond to the singlets
$\overline{Q}$, $Q$ and $H$ respectively,
and, as in the electric theory, there is no superpotential for
the light fields.

The $SU(N)$ theory with an antisymmetric tensor
studied in this letter is a chiral theory and is thus
susceptible to break supersymmetry dynamically
\refs{\ads, \mur,\pop}.
The question of SUSY
breaking is analyzed here {}from the point of view of the
 magnetic theory. One considers the
theory for $N$ odd and $F=5$. One first integrates
out one flavor of $Q$ and $\overline{Q}$
to leave the non-Abelian Coulomb
phase. The magnetic theory is completely higgsed
and very weakly coupled. It consists of the
fields $M$, $H$, $B_1$, $B_3$ and $\overline{B}$ and
a superpotential proportional to
$B_1M^3H^{(N-3)/2}+ B_3 M H^{(N-1)/2}+\overline{B}B_1B_3$
which describes the theory everywhere on the moduli space.
The last term comes {}from the
term $B_1 p\overline{q}$ of the magnetic
superpotential. We included the first two terms
to agree with the results described earlier,
even though we do not know the actual
origin of these terms.
{}From this $F=4$ theory, it is straightforward to show
that the theory with $F=0$ breaks SUSY dynamically.
We add a rank 4 mass term $\sum_{i,j=1}^4 m^{ij}Q_i\overline{Q}_j$ and
Yukawa couplings $\lambda A\overline{QQ}$ over all the
$\overline{Q}$s except $\overline{Q}_i$, $i=1,\ldots,4$
(i.e. $\lambda$ has rank $N-1$).
This lifts all the classical flat directions of the electric theory.
Now study the effect of these perturbations on the dual.
First $B_1B_3=0$ by the equation of motion for $\overline{B}$. Then,
multiplying the $M$ equation of motion by $B_1$, we get that $B_1=0$;
thus $m=H^{k-1} B_3$; multiplying this
equation by $B_3$, it says that $B_3=0$,
implying $m=0$,
a contradiction. Thus
dynamical supersymmetry breaking occurs at tree level in the
dual description.

Related results about SUSY breaking in the dual description
are discussed in \refs{\ithomas, \poustr}.

\centerline{\bf Acknowledgments}

We would like to thank M. Berkooz, D. Kutasov,
R. Leigh, N. Seiberg, A. Schwimmer and
S. Thomas for useful discussions.  This
work was supported in part by DOE grant \#DE-FG05-90ER40559
and by a Canadian 1967 Science fellowship.

\listrefs
\bye